%% file: main.tex
\documentclass[showpacs,amsmath,amssymb,twocolumn,superscriptaddress,notitlepage,preprintnumbers,aps,pra,longbibliography]{revtex4-1}

\usepackage{amsmath}
\usepackage{bbm}
\usepackage[normalem]{ulem}
\usepackage{times}
\usepackage[normalem]{ulem}
\usepackage{graphicx}
\usepackage{array}
\newcolumntype{P}[1]{>{\centering\arraybackslash}p{#1}}
\usepackage{dcolumn} 
\usepackage{bm} 
\usepackage{tabularx}
\usepackage{enumerate}
\usepackage{multirow}
\usepackage{float}
\usepackage{amsthm}
\usepackage{amsfonts,amssymb}
\usepackage{mathrsfs}
\usepackage{subfigure}
\usepackage{pifont}
\usepackage{thmtools}
\usepackage{verbatim}
\usepackage[ruled,vlined]{algorithm2e}

\usepackage{xcolor}

\usepackage{qcircuit}

\usepackage{hyperref}
\usepackage{varioref}
\usepackage{soul}
\hypersetup{colorlinks=true, linkcolor=blue, citecolor=blue, urlcolor=blue}

\input{macro_style}

\input{macro_quantum}

\input{macro_math}

\input{macro_type}
\newcommand{\RV}{\widetilde}

\newcommand{\utchem}{Department of Chemistry, University of Toronto, Toronto, Ontario M5G 1Z8, Canada}
\newcommand{\utcomp}{Department of Computer Science, University of Toronto, Toronto, Ontario M5S 2E4, Canada}
\newcommand{\vectorinst}{Vector Institute for Artificial Intelligence, Toronto, Ontario M5S 1M1, Canada}
\newcommand{\cifar}{Lebovic Fellow, Canadian Institute for Advanced Research, Toronto, Ontario M5G 1Z8, Canada}
\newcommand{\material}{Department of Materials Science \& Engineering, University of Toronto, Toronto, Ontario M5S 3E4, Canada}
\newcommand{\chemical}{Department of Chemical Engineering \& Applied Chemistry, University of Toronto, Toronto, Ontario M5S 3E5, Canada}
\newcommand{\aist}{
    Research Center for Emerging Computing Technologies, National Institute of Advanced Industrial Science and Technology (AIST), 1-1-1 Umezono, Tsukuba, Ibaraki 305-8568, Japan}
\newcommand{\keio}{
    Quantum Computing Center, Keio University, 3-14-1 Hiyoshi, Kohoku-ku, Yokohama, Kanagawa, 223-8522, Japan}

\begin{abstract}
    Estimation of the expectation value of observables is a key subroutine in quantum computing and is also the bottleneck of the performance of many near-term quantum algorithms. Many methods have been proposed to reduce the number of measurements needed for this task by designing measurement schemes that decide the measurements to perform; however, these schemes are usually constructed from hand-crafted heuristics, which limits the measurement efficiency they can achieve.
    In this paper, we propose a framework for learning measurement schemes directly from the observable, using machine learning techniques including stochastic gradient descent and a two time-scale update rule.
    As a concrete realization of this framework, we introduce Composite-Locally Biased Classical Shadow (C-LBCS), which learns a mixture of locally-biased classical shadows and their mixing weights end-to-end.
    We numerically demonstrate C-LBCS on molecular systems up to $\mathrm{CO}_2$ (30 qubits) and show that C-LBCS outperforms the previous state-of-the-art methods despite its simplicity.
    We believe our approach opens up a reliable and scalable path toward efficient observable estimation on large quantum systems.
\end{abstract}

\begin{document}
\title{Machine learning of measurement schemes for efficient quantum observable estimation}
\author{Zi-Jian Zhang}
\affiliation{\utcomp}
\affiliation{\vectorinst}
\author{Kouhei Nakaji}
\affiliation{\utchem}
\affiliation{\aist}
\affiliation{\keio}
\author{Matthew Choi}
\affiliation{\utcomp}
\affiliation{\vectorinst}
\author{Al\'an Aspuru-Guzik}
\email{aspuru@utoronto.ca}
\affiliation{\utchem}
\affiliation{\utcomp}
\affiliation{\vectorinst}
\affiliation{\material}
\affiliation{\chemical}
\affiliation{\cifar}
\date{\today}
\maketitle


\section{Introduction}
Quantum technology makes it possible to create and measure entangled quantum states living in high-dimensional Hilbert spaces, enabling the implementation of quantum algorithms fundamentally faster than their classical counterparts~\cite{arute2019quantum,zhong2020quantum,madsen2022quantum}. In these algorithms, the measurement of quantum states plays a critical role. On one hand, it converts intangible quantum states to classical results that can be recognized and serve as outputs. On the other hand, the destructive nature of quantum measurements forces one to prepare the quantum states multiple times, forming the bottleneck of many quantum applications. Significantly, the emergence of noisy intermediate-scale quantum (NISQ) devices~\cite{preskill2018quantum, bharti2022noisy} puts efficient measurement methods in increasing importance since near-term quantum algorithms typically involve estimating the expectation value of complicated observables~\cite{bharti2022noisy,cerezo2021variational,peruzzo2014variational, Li2017, stair2020multireference, grimsley2019adaptive, zhang2023low}. 
There are also error mitigation methods~\cite{doi:10.7566/JPSJ.90.032001, Li2017, PhysRevX.11.041036,sun2020mitigating} proposed for trading the number of measurements with the accuracy of the result, making measurement methods a significant factor in the overall performance of near-term algorithms.

The problem of estimating the expectation value of observables on a quantum state can be formulated as follows: there is an observable $O=\sum_{j}a_jO_j$, where $\vec{a}$ are real coefficients and $\{O_j\}$ are easily measurable fragments of the observable $O$. Given a quantum state $\ket{\psi}$ that we can prepare, the problem is how to estimate the expectation value $\langle O \rangle = \bra{\psi}O\ket{\psi}$ to a certain accuracy with fewer copies (shots) of $\ket{\psi}$. We note that for quantum computing with near-term quantum devices, we usually limit the family of measurements to Pauli measurements so that they can be easily implemented; correspondingly, the observable is decomposed into a sum of the tensor product of Pauli operators (Pauli strings) as $O = \sum_j a_j P_j$. 
If one naively estimates the expectation value by independently performing the estimation on each Pauli string, to a fixed accuracy, the required number of measurements scales quadratically with $\|\vec{a}\|_1$. This will be problematic when larger systems of practical interest are considered.

Various measurement methods have been proposed to mitigate the problem above and there are roughly two important families of them. The first family outputs a probability distribution on a small set of measurements for a given observable. 
The measurements to be made can then be sampled from the set. The methods in the family can be represented by the largest degree first (LDF) grouping~\cite{verteletskyi2020measurement}, overlapped group measurement (OGM)~\cite{wu2023overlapped}, and other recent approaches~\cite{yen2023deterministic,choi2022improving}. 
An additional large family of methods is characterized by
using ideas from
classical shadow (CS)~\cite{huang2020predicting} and locally-biased classical shadow (LBCS)~\cite{hadfield2022measurements}. These methods are not associated with a small set of measurements to sample from. Instead, they employ distributions on all Pauli measurements.
This family of methods provides a feasible way to measure the observables even when they contain exponentially many terms~\cite{huang2020predicting}. 
However, these methods usually only offer distributions of measurements with a simple structure and cannot provide the best performance for molecular Hamiltonians~\cite{wu2023overlapped}. This drawback makes them currently not the best choice for applications such as variational quantum eigensolver (VQE)~\cite{peruzzo2014variational}. Therefore, how to enhance the representation power of these methods so that they have a better shot efficiency for complicated observables, becomes an interesting question.

In this work, we introduce a framework for learning measurement schemes directly from the observable. Rather than relying on hand-crafted heuristics, we parameterize a measurement schemes and train them by minimizing a tractable cost function using machine learning techniques. The key technical building block is the composite measurement scheme (CMS), which combines multiple base measurement schemes via a trainable mixing weights.
As a concrete instance, we introduce composite locally-biased classical shadow (C-LBCS), which learns an optimized mixture of LBCS schemes.
We numerically show that by jointly optimizing the mixing weights and the parameters of each sub-scheme with a gradient rescaling strategy and a two time-scale update rule (TTUR)~\cite{heusel2017gans}, our learned measurement scheme achieves state-of-the-art measurement efficiency, even when stochastic gradient descent with a small batch size is adopted.

The rest of the paper is organized as follows. In \autoref{section:framework}, we present the learning framework and the composite measurement scheme as its parameterization.
Then, in \autoref{section:optimization}, we propose several optimization strategies tailored to training measurement schemes. In \autoref{section:numerics}, we demonstrate C-LBCS and compare its performance against previous methods on various molecular Hamiltonians.
Finally, in \autoref{section:discussion}, we conclude with discussion and a general outlook.
\section{Framework}
\label{section:framework}
\subsection{Measurement schemes}
In this work, we define the term \textit{measurement schemes} (MS) as measurement generators that input the required number of measurements and output a set of measurements to perform.

\begin{definition}[Measurement scheme]
A measurement scheme $S$ is a generator of measurements that outputs a multi-set of measurements on the objective quantum system given the required number of measurements (shot number) and optionally other information.
\end{definition}

A measurement scheme may contain optimizable parameters. For example, in the LBCS method \cite{hadfield2022measurements}, the probabilities of generating each Pauli operator on each qubit are the parameters to be optimized.
In OGM~\cite{wu2023overlapped}, the parameters are the probabilities of generating each Pauli measurement (group) that is constructed by the method. We note that we formulate these optimizable parameters as included within a measurement scheme rather than provided as inputs.

Here, we emphasize the difference between the term \textit{measurement scheme} and the term \textit{measurement method}. 
By \textit{measurement method}, we imply it is an end-to-end protocol for estimating the expectation value of observables, which includes the optimization of the parameters, the generation of measurements, the measurement process and the post-processing protocol that synthesizes the final result from measurement outcomes. In this sense, previously proposed methods such as OGM \cite{wu2023overlapped} and LBCS \cite{hadfield2022measurements} can be recognized as measurement methods that contain a measurement scheme. Measurement schemes, on the other hand, are just generators of measurements. Based on this definition, there is a special family of measurement schemes that has a memory-less (Markovian) structure.

\begin{definition}[Simple measurement scheme]
A simple measurement scheme $S$ is a measurement scheme that can be implemented by sampling measurements from a fixed distribution for each required measurement.
\end{definition}

A simple measurement scheme only inputs the number of measurements needed and the generation of each measurement is independent. Many proposed measurement schemes (e.g. the ones used in OGM and LBCS) can be modelled by simple measurement schemes and their simple structure makes them easy to analyze and serve as a base for generalization.

\begin{figure}[t]
    \centering
    \includegraphics{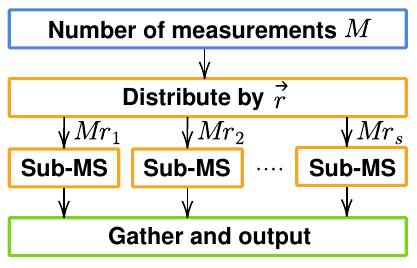}
    \caption{A diagrammatic representation of the composite measurement scheme defined in \autoref{def:sample-prod}. The total number of measurements $M$ is distributed to sub-measurement-schemes (Sub-MS) by the probabilities $\vec{r}$. The composite scheme gathers the measurements generated by all the sub-schemes as the output.}
    \label{fig:cms-diagram}
\end{figure}

\subsection{Composite measurement schemes}

Next, we introduce a way to combine multiple simple measurement schemes to make a CMS.
Suppose $\vec{S}$ is a list of schemes to be combined. A natural way to combine multiple schemes is to assign each scheme a probability; when generating a measurement from the combined scheme, we sample a scheme according to the probability and generate a measurement from the sampled scheme (see \autoref{fig:cms-diagram} for a diagrammatic representation). Specifically, we propose the $\mathrm{SampleProd}$ operator to represent this combination.

\begin{definition}[Composite measurement scheme]
Suppose there is a list of simple measurement schemes $\vec{S}$ and there is a distribution of the schemes in $\vec{S}$ represented by the probabilities $\vec{r}$, in which $r_k$ corresponds to $S_k$. We define $S' = \mathrm{SampleProd}(\vec{r}, \vec{S})$ to be the composite measurement scheme which generates each measurement by first sampling a sub-scheme $S_{k'}$ from $\vec{S}$ by the probabilities $\vec{r}$ and then generating a measurement from $S_{k'}$.
\label{def:sample-prod}
\end{definition}

CMS can be used to form interesting new types of measurement schemes. For example, it is viable to combine shadow methods with grouping methods by making $\mathrm{SampleProd}$ of their measurement schemes.

More importantly, CMS provides a way to scale up measurement schemes. In this work, we will focus on demonstrating $\mathrm{SampleProd}$ by applying it to LBCS schemes as an example of CMS. The measurement scheme of LBCS has a distribution $\beta_i$ on the set of Pauli operators ($X, Y$ and $Z$) for each qubit $i$ in the system. When generating a measurement, for the $i$-th qubit, the scheme samples a Pauli operator on the qubit by the distribution $\beta_i$. The Pauli measurement generated will be the tensor product of the Pauli operators sampled for each qubit. Suppose $\vec{S}$ is a list of LBCS schemes, $\operatorname{SampleProd}(\vec{r}, \vec{S})$ offers a natural generalization of single LBCS schemes; this is the C-LBCS method studied in this work.

\begin{algorithm}
 \caption{Composite locally-biased classical shadow (C-LBCS)}
 \begin{enumerate}
     \item Suppose the composite scheme is made by $\operatorname{SampleProd}(\vec{r}, \vec{S})$, in which each sub-scheme $S_k$ is a LBCS scheme with adjustable distributions $\{\beta^{k}_i\}$.
     \item When $M$ measurements are required, repeat the following process $M$ times.
     \begin{itemize}
        \item Sample a sub-scheme $S_{k'}$ by the probabilities $\vec{r}$. Initialize a Pauli string $Q$ with $Q[i]=I$ for all $i$.
        \item For each qubit $i$, decide the Pauli operator $Q[i]$ on it by sampling a Pauli operator $P$ from $\beta_{i}^{k'}$ and setting $Q[i]$ to $P$.
        \item Output $Q$ as the Pauli measurement to carry out.
     \end{itemize}
 \end{enumerate}
\end{algorithm}

C-LBCS has more parameters and stronger representation power than LBCS. Moreover, the representation power of a C-LBCS scheme can be adjusted by the number of sub-schemes. In the extreme case when there is only one sub-scheme, C-LBCS will degenerate to the original LBCS. 

C-LBCS differs from previous methods by providing a \textit{top-down} approach to improve the measurement efficiency, whereas most of the previous methods use the \textit{bottom-up} approach. Previous methods usually solve the problem by considering how to improve a certain measurement scheme (the \textit{bottom}) by certain heuristics. For example, in grouping methods, the scheme is usually $l_1$ sampling and commuting relations are leveraged to improve it. In the Derand method~\cite{PhysRevLett.127.030503}, one improves the distribution from the classical shadow by greedily optimizing a cost function. These methods usually do not involve the concept of an optimal scheme in their derivation since there is no direct way to involve it in its heuristics. However, in C-LBCS, we can directly consider how to approximate an optimal simple scheme (the \textit{top}) and provide heuristics in a top-down manner. 

In the following, we show that C-LBCS is capable of representing any distribution of Pauli measurements that are applied to the whole system.

\begin{theorem}[Universality of C-LBCS]
\label{thm:universality}
Suppose there are $n_q$ qubits in a system. When there are $3^{n_q}$ sub-schemes, a C-LBCS can simulate any distribution of Pauli measurements if every Pauli measurement $Q_k$ in the distribution acts non-trivially on every qubit.
\end{theorem}
\begin{proof}
Denote the set of Pauli measurements in the distribution and their probability to be simulated by $\vec{Q}$ and $\vec{p}$. To simulate this distribution, one just needs to set sub-scheme $S_k$ in the C-LBCS scheme to output $Q_k$ and set $\vec{r}=\vec{p}$. Notice that there are at most $3^{n_q}$ different Pauli measurements that act on every qubit in the system. Therefore, at most $3^{n_q}$ sub-schemes are needed.
\end{proof}

This universality result should be understood as an expressivity statement rather than a claim that exponentially many sub-schemes are practical or optimizable.
Though C-LBCS might need exponentially many sub-schemes to simulate the optimal simple scheme, in \autoref{sec:var-vs-method}, we show that state-of-the-art efficiency can be provided by C-LBCS with as many sub-schemes as groups in the OGM method. This implies the number of sub-schemes we need to achieve good efficiency can be far less than exponentially large. In practice, one can adjust the number of sub-schemes by the available computational resources and the required efficiency.

\section{Optimization of CMS}
\label{section:optimization}
One problem with CMS is how to determine the probability of each sub-scheme and calculate its optimal parameters. In the following, we discuss how to optimize CMS, including both the parameters of the sub-schemes and their probabilities, with the assumption that all the measurements involved are Pauli measurements and the state to be measured is totally unknown.
We introduce a cost function which has a simple physical interpretation and can be constructed without the knowledge of the target quantum states. Then, we analyze the structure of the cost function and investigate how to perform gradient descent with it. 

\subsection{Average one-shot variance}

A straightforward way to quantify the performance of a measurement method is using the variance of the estimation it produces, given a fixed number of shots. However, as we mentioned, a post-processing protocol must be specified for measurement schemes producing estimations.
In the following, we formalize and adopt a post-processing protocol that is widely adopted~\cite{shlosberg2023adaptive, yen2023deterministic}, in which estimations of terms in the observable are generated and summed up to the estimation of the whole observable. This post-processing protocol is different from the one adopted in Ref.~\cite{hadfield2022measurements, huang2020predicting}, in which only estimations to the whole observable are generated and averaged. 

We note that throughout this work, we use $\RV{\cdots}$ to represent random variables. Suppose there is a list of Pauli measurement $\{Q_k\}$ generated from the measurement scheme and $\{\RV{x}_k\}$ are the corresponding results in the form of bitstrings. Also, we define the following relation that characterizes what Pauli measurements can produce estimations for the expectation value of a Pauli string.
\begin{definition}[Qubit-wise covering]
For a Pauli string $P$ and a Pauli measurement represented by the Pauli string $Q$, we say $Q$ covers $P$, or equivalently $P\triangleright Q$, when $P[i]$ equals to either $Q[i]$ or identity for all qubit $i$.
\end{definition} 

With the above notations, we present \autoref{alg:estimate}, which is set to be the post-processing protocol for demonstrating all measurement schemes in this work.

\newcommand{\compat}{\triangleright}
\begin{algorithm}[h]
    \caption{Observable estimation by estimating each term \label{alg:estimate}}
    \begin{itemize}
        \item For each term $P_j$ in the observable $O=\sum_j a_j P_j$, generate an estimation $\RV{\expectv{P_j}}$ for it by averaging one-shot estimations
        \begin{equation}
            \RV{\expectv{P_j}} := \frac{1}{m_j}\sum_{k,P_j \triangleright Q_k} \mu(P_j, \RV{x}_k),
        \end{equation}
        where $m_j = \sum_{k,P_j \triangleright Q_k} 1$ is the number of available one-shot estimations
        and 
        \begin{equation}
                \mu(P_j, \RV{x}_k) = \prod_{i, P_{j}[i]\neq I} (-1)^{\RV{x}_{k}[i]},
            \end{equation}
        is the one-shot estimation generated by each measurement outcome $\RV{x}_k$. Here, $\RV{x}_{k}[i]$ is used to denote the $i$-th bit of $\RV{x}_k$.

        \item Output the estimation of $\expectv{O}$ by summing up the estimations of each term.
        \begin{equation}
            \RV{\expectv{O}} = \sum_j a_j \RV{\expectv{P_j}}.
        \end{equation}
    \end{itemize}
\end{algorithm}

For a list of Pauli measurements $\{Q_k\}$, the output $\RV{\expectv{O}}$ of \autoref{alg:estimate} is random due to the random nature of quantum mechanics. The variance of the estimate of $\RV{\expectv{O}}$ can be calculated by (See \appref{sec:var-expression} for derivation)
\begin{equation}
\label{equ:exact-variance}
    \begin{split}
        \Var(\RV{\expectv{O}}) &= \sum_{j,\ell} a_j a_{\ell} \Cov(\RV{\expectv{P_j}}, \RV{\expectv{P_{\ell}}})\\
&= \sum_{j,\ell} a_j a_{\ell}  \frac{m_{j\ell}}{m_{jj}m_{\ell\ell}} (\expectv{P_j P_{\ell}} - \expectv{P_j}\expectv{P_{\ell}}),
    \end{split}
\end{equation}
where 
\begin{equation}
	m_{j\ell} := \sum_{k, P_j \triangleright Q_k, P_\ell \triangleright Q_k} 1,
\end{equation}
and $\expectv{\cdots}$ here are the true expectation values that depend on the state measured.
$m_{jj}$ equals $m_j$ that we defined above and $m_{j\ell}\;(j\neq \ell)$ is the number measurements that generate a one-shot estimation for both $\expectv{P_j}$ and $\expectv{P_{\ell}}$.

When the measurement scheme is stochastic, $m_{j\ell}$ needs to be modelled as a random variable $\RV{m}_{j\ell}$ and the variance then needs to be obtained by applying the law of total variance.
\begin{align}
&\Var(\RV{\expectv{O}})
= \sum_{j,\ell} a_j a_{\ell}  \ExpV{\frac{\RV{m}_{j\ell}}{\RV{m}_{jj}\RV{m}_{\ell\ell}}} (\expectv{P_j P_{\ell}} - \expectv{P_j}\expectv{P_{\ell}}), 
\end{align}
The above equation involves $\expectv{P_j}$ and $\expectv{P_j P_{\ell}}$ which depends on the state that is measured. As we assume the target state is unknown, it cannot be directly used as the cost function.
Therefore, we propose to use the variance averaged over the whole Hilbert space with the Haar measure using the following lemma.
\begin{lemma}
Suppose there are $n_q$ qubits in the system, then
\label{thm:average_cov}
\begin{equation}
    \int_{\mathrm{Haar}} \Big(\expectv{P_j P_{\ell}} - \expectv{P_j}\expectv{P_{\ell}}\Big) \mathrm{d}\ket{\psi}= \delta_{j\ell} \frac{2^{n_q}}{2^{n_q} + 1},
\end{equation}
where $\delta_{j\ell}$ is the Kronecker delta function.
\end{lemma}
A proof of the lemma is given in \appref{section:proofLemmaHaar}. With \autoref{thm:average_cov}, we have
\begin{equation}
\label{eq:varHaarAverage}
    \Exp_{\ket{\psi}\sim\mathrm{Haar}}[\Var[\RV{\bra{\psi}{O}\ket{\psi}}]] = \sum_{j} a_j^2  \ExpV{\frac{1}{\RV{m}_{j}}} \frac{2^{n_q}}{2^{n_q}+1},
\end{equation}
which is independent of the state to be measured. We remark that we can also show that the variance of the variance $\Var(\RV{\expectv{O}})$ over the whole Hilbert space with the Haar measure is exponentially suppressed (see Section~4 in \cite{nakaji2022measurement}). Therefore, $\Var(\RV{\expectv{O}})$ tends to take the average value \autoref{eq:varHaarAverage} when sampling $|\psi\rangle$ according to the Haar measure. 

Notice that $\ExpV{\frac{1}{\RV{m}_{j}}}$ in the right-hand side of \autoref{eq:varHaarAverage} is ill-defined when the probability that $\RV{m}_{j}$ equals $0$ is non-zero, which is usual when $M$ is finite. 
To make a well-defined quantifier, we define the ratio between $\Exp_{\ket{\psi}\sim\mathrm{Haar}}[\Var[\RV{\bra{\psi}{O}\ket{\psi}}]]$ and $1/M$ in the limit that $M\rightarrow +\infty$:
\begin{align}
    V &= \lim_{M\rightarrow +\infty} \sum_{j} a_j^2  
    \ExpV{\frac{M}{\RV{m}_{j}+\epsilon}} \frac{2^{n_q}}{2^{n_q}+1},
\end{align}
in which the random variable $\frac{M}{\RV{m}_{j}+\epsilon}$ will converge to $\frac{M}{\RV{m}_{j}}$ when $M\rightarrow +\infty$ if we set $\epsilon\in o(M)$. In \appref{sec:probability}, we show that if $\epsilon$ is a positive number in $\Theta(M^{\frac{5}{6}})$ and $\RV{m}_{j}$ is a well-behaved random variable whose expectation value and variance are both in $\Theta(M)$, we have
\begin{align}
&\lim_{M \rightarrow +\infty}\ExpV{\frac{M}{\RV{m}_{j}+\epsilon}} = \lim_{M \rightarrow +\infty} \frac{M}{\ExpV{\RV{m}_{j}}}=\frac{1}{h_j},
\end{align}
where $h_j$ can be interpreted as the average probability for the term $P_j$ being covered by a measurement generated by the scheme. As long as $h_j>0$ for all $j$, $V$ is well-defined as expected. When $V$ is instead evaluated empirically from a finite set of measurements, some terms may be left uncovered ($h_j = 0$); we describe how such terms are handled in \appref{sec:uncovered-terms}. This formula holds for all simple measurement schemes, since in that case, $\RV{m}_{j}$ obeys the binomial distribution whose expectation value and variance grow linearly with $M$ (therefore in $\Theta(M)$). Therefore, in the following, we can reasonably define the cost function we use in this work.

\begin{definition}[Average one-shot variance]
Suppose there is a $n_q$ qubit observable $O=\sum_j a_jP_j$ and a simple measurement scheme $S$. The average one-shot variance $V$ of $S$ for $O$ is
\begin{equation}
     V = \sum_{j} \frac{a_j^2}{h_{j}} \frac{2^{n_q}}{2^{n_q}+1},
\end{equation}
where $h_j=\lim_{M\rightarrow\infty} \ExpV{\RV{m}_j} / M$.
\end{definition}

The physical meaning of $V$ is the scaling factor of the variance of $\RV{\expectv{O}}$ averaged by Haar measure in the limit as the number of measurements tends to infinity. A scheme can be optimized by this cost function as long as $\vec{h}$ can be efficiently estimated given the parameters of the scheme. 
Notably, the structure of $\operatorname{SampleProd}$ allows a further decomposition of $\vec{h}$ by the contribution of each sub-scheme.
Define $\RV{m}_j^k(M_k)$ to be the number of measurements that cover $P_j$ generated by the sub-scheme $S_k$ with $M_k$ measurements assigned to it.
Denote $h_j^k=\lim_{M_k \rightarrow+\infty} \ExpV{\RV{m}_j^k(M_k)} / M_k$. $h_j$ can be decomposed as
\begin{align}
    h_j &= \lim_{M \rightarrow +\infty} \sum_{k} \ExpV{\RV{m}_j^k(M_k)}/M  \\
    &= \sum_{k} \lim_{M \rightarrow +\infty} (\ExpV{\RV{m}_j^k(M_k)}/M_k) (M_k/M) \\
    &= \sum_{k} r_k h_j^k,
\end{align}
where we used that $\lim_{M\rightarrow +\infty} M_k/M = r_k$.
In this way, $V$ can be rewritten as 
\begin{equation}
\label{equ:cost-fun-decomposed}
    V 
    = \frac{2^{n_q}}{2^{n_q}+1} 
    \sum_j \frac{a_{j}^2}{\sum_{k} r_k h^k_j},
\end{equation}
which will be adopted in all the following sections.
In the case of C-LBCS, $h^k_j$ is simply the probability that $P_j$ is covered by the sampled measurement when the $k$-th LBCS scheme has been sampled. We put the detail of its calculation in \appref{sec:h-for-c-lbcs}.

\subsection{Optimization strategy}
After setting the cost function, we discuss how to efficiently optimize a CMS made by $\operatorname{SampleProd}$. In $V$, there are two sets of parameters to be optimized, the probabilities $\vec{r}$ and the parameters $\{\theta_i^k\}$ for each sub-scheme. Here, $\theta_i^k$ denotes the $i$-th parameter for the sub-scheme $S_k$; in C-LBCS, the raw parameters $\{\theta_i^k\}$ are mapped to the per-qubit LBCS distributions $\{\beta_i^k\}$ by applying $\operatorname{SoftPlus}$ element-wise followed by normalization (see \appref{section:detailTraning}), and these distributions determine the coverage probabilities $\{h_j^k\}$ appearing in \autoref{equ:cost-fun-decomposed}.

\subsubsection{Gradient rescale}
A straightforward way to optimize $\vec{r}$ and $\{\theta_i^k\}$ is to apply a gradient descent directly. To this end, we calculate the gradient on the parameter $\theta^k_i$ as
\begin{align}
\label{eq:vGradient}
    \frac{\partial V}{\partial \theta^k_i} = \sum_j
    \frac{\partial V}{\partial h_j} \frac{\partial \sum_k r_k h^k_j}{\partial \theta^k_i} =
    \sum_j \frac{\partial V}{\partial h_j}r_k\frac{\partial h^k_j}{\partial \theta^k_i}.
\end{align}
The gradient on $\theta^k_i$ 
can be very small when $r_k$ is close to zero, which may happen when parameters in $S_k$
are poorly initialized
and $r_k$ is optimized to nearly zero for avoiding $S_k$ being sampled. 
In this way, $S_k$ will be frozen out from the composite scheme as its parameters stop updating and its probability to be sampled is nearly zero.
This situation should be avoided as we want to utilize all sub-schemes. 
Thus, we adopt a strategy to rescale the gradient by $1/r_k$ when optimizing the parameters of $S_k$. Specifically, we use 
\begin{equation}
\frac{1}{r_k} \frac{\partial V}{\partial \theta^k_i}
=    \sum_j  \frac{\partial V}{\partial h_j} \frac{\partial h^k_j}{\partial \theta^k_i}.
\end{equation}
instead of \autoref{eq:vGradient} in the gradient descent.
The rescaled gradient ensures that each sub-scheme continues to be optimized even if their $r_k$ is small.

\subsubsection{Two time-scale update rule}

As the second technique, we introduce TTUR~\cite{heusel2017gans}, in which the parameters of the model are divided into two parts and optimized with different learning rates. Before going into detail, we clarify the feature of the cost function \autoref{equ:cost-fun-decomposed} in terms of convexity.
As the term $r_k h^k_j$ is included, the cost function \autoref{equ:cost-fun-decomposed} is non-convex, and the optimization may be trapped in a local minimum. However, we show that the cost function $V$ is convex as a function of $\vec{r}$ (with $\{\theta^k_i\}$ and therefore $\{h^k_j\}$ fixed) or $\{h^k_j\}$ (with $\vec{r}$ fixed).

The key observation is that the denominator $\sum_{k} r_k h^k_j$ in each term of \autoref{equ:cost-fun-decomposed} is affine in $\{h^k_j\}$ when $\vec{r}$ is fixed, and affine in $\vec{r}$ when $\{h^k_j\}$ is fixed. Since $f(x)=1/x$ is convex for $x>0$ and the composition of a convex function with an affine map preserves convexity, each term $a_j^2/(\sum_{k} r_k h^k_j)$ is convex in either set of variables separately, and so is $V$ as a summation of these terms with non-negative weights. A formal statement and proof of this biconvexity are given in \appref{section:appendix-biconvex}.

This bi-convex structure (proved in \appref{section:appendix-biconvex}) makes the cost function substantially easier to optimize than the highly non-convex loss landscapes typical of neural networks~\cite{keskar2016large}: each subproblem is convex when the other set of parameters is fixed. We emphasize that the biconvexity holds in the variables $(\vec{r}, \{h^k_j\})$ and does not extend to the underlying sub-scheme parameters $\{\theta^k_i\}$: in C-LBCS, each $h^k_j$ is a product of per-qubit probabilities (\appref{sec:h-for-c-lbcs}), and this nonlinear map breaks convexity, so the subproblem over $\{\theta^k_i\}$ with $\vec{r}$ fixed is in general non-convex (see \appref{section:appendix-biconvex}). Nevertheless, the convexity of $V$ when $\vec{r}$ or $\{h^k_j\}$ is fixed suggests that the optimization is better behaved when the two sets of parameters are updated separately. This situation is similar to the training of a generative adversarial network (GAN)~\cite{GAN}, in which the training can also be divided into two parts (the generator and the discriminator), whose optimization is easier with the parameters of the other part fixed. Therefore, we propose to adopt the same strategy in the optimization of the probabilities $\vec{r}$ and parameters in the sub-schemes  and set different learning rates for $\vec{r}$ and $\{\theta^k_i\}$ when optimizing a CMS. Throughout this work, we will set the learning rate of $\vec{r}$ to be 10 times smaller than the learning rate of $\{\theta^k_i\}$.

\subsubsection{Stochastic gradient descent}

Stochastic gradient descent (SGD) is a widely adopted strategy in machine learning, in which the gradient provided to the optimizer is generated only with part of the whole dataset. SGD provides a general approach for training models when the dataset is large. In our case, the models (measurement schemes) are trained with the terms from the observable. An observable might contain a large number of terms in near-term quantum algorithms. For electronic structure problems (the main target of VQE), the Hamiltonian contains $\mathcal{O}(n_O^4)$ terms~\cite{RevModPhys.92.015003}, with $n_O$ being the number of spin-orbitals. 
If the Hamiltonian is not split, it will be hard to fit the computation in the memory of a GPU.

Therefore, we propose to use SGD in the training of measurement schemes. Specifically, in each epoch, we randomly split the terms of observable $O$ into batches $\{B_l\}$, with each of them containing a nearly fixed number of terms and $\sum_l B_l=O$. Then, we iterate over $\{B_l\}$ and calculate the gradient in each step with one batch. After iterating all the batches, we split $O$ with a different seed for the next epoch.
We show in \autoref{sec:training} that C-LBCS can perform well even when the batch size is very small.

\section{Numerical examples for C-LBCS}
\label{section:numerics}

In the following, we numerically demonstrate the performance of C-LBCS with the average one-shot variance as the quantifier.
In C-LBCS, the number of sub-schemes affects the performance of the composite scheme. We show that a C-LBCS can outperform previous state-of-the-art methods by using a proper number of sub-schemes. We also analyze how different training strategies can affect the convergence of optimization. 

Throughout this section, by default, both the Rescale and TTUR strategies are adopted in the training. 
All probabilities ($\vec{r}$ and $\{\beta^k_i\}$) are calculated by passing parameters into $\operatorname{SoftPlus}$ and normalization layers (See \appref{section:detailTraning}).
The batch size for stochastic gradient descent is set to $500$ Pauli strings and the learning rate for all optimizations is $5\times 10^{-3}$ for sub-scheme parameters and $5\times 10^{-4}$ for $\vec{r}$. Optimizations terminate when the cost function fails to decrease more than $0.1\%$ in the past $1000$ steps. 
The initial parameters for the LBCS schemes are set so that the C-LBCS scheme resembles the $l_1$ sampling scheme. For a C-LBCS scheme with $n_S$ sub-schemes, we select the $n_S$ terms with the largest weights ($|a_j|$) in $O$ and set the LBCS sub-schemes to generate them initially. The probabilities $\vec{r}$ of each sub-scheme are set to be proportional to the weight of those terms correspondingly.

The molecular Hamiltonians are generated by mapping the Fermionic Hamiltonian for the electronic structure problem with Jordan-Wigner~(JW)~\cite{wigner1928paulische} and Bravyi-Kitaev~(BK)~\cite{bravyi2002fermionic} transformations. All molecules are in their equilibrium geometry and under the STO-3G basis set, except for hydrogen chains, in which the hydrogens are spaced by $2$ Bohr radius and the STO-6G basis set is used. The equilibrium geometries are retrieved from \cite{johnson2022nist}.

\subsection{Performance}
\label{sec:var-vs-method}

\begin{figure*}[t]
    \centering
    \includegraphics[width=0.9\linewidth]{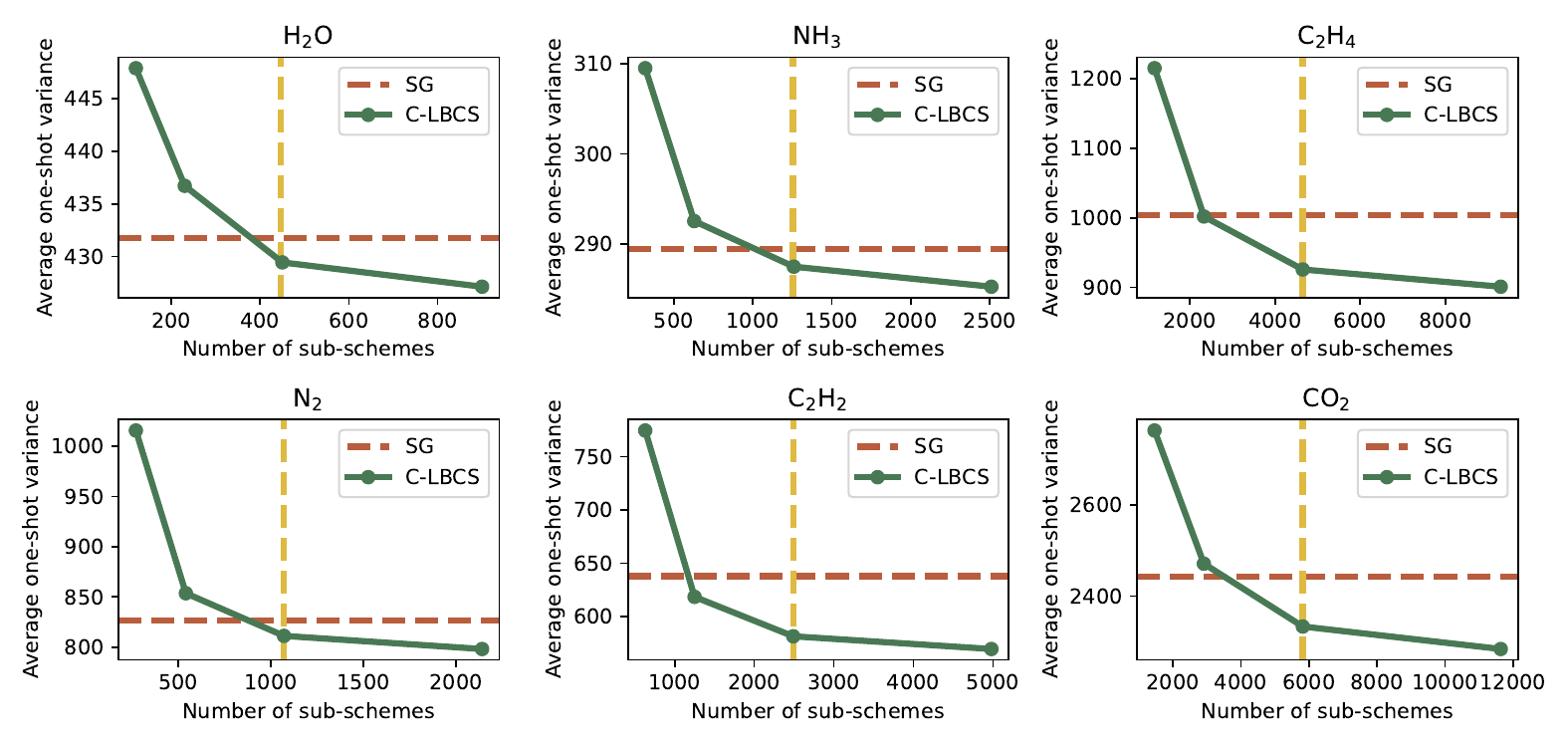}
    \caption{The performance of C-LBCS with different numbers of sub-schemes, quantified by the average one-shot variance. All the Hamiltonians are encoded by JW transformation. The vertical yellow line represents the number of groups produced by the OGM method. The horizontal red line represents the average one-shot variance of the ShadowGrouping method.
    }
    \label{fig:subscheme-number}
\end{figure*}

We test our method by comparing its performance on molecular Hamiltonians against previous methods, including the derandomized classical shadow (Derand)~\cite{PhysRevLett.127.030503}, OGM and ShadowGrouping (SG) \cite{gresch2023guaranteed}.
We note that SG relies on a greedy algorithm to construct its measurement scheme: at each step it selects the next measurement greedily to maximize coverage, making the scheme construction inherently sequential. Although the candidate scoring within each step can be parallelized, the outer loop of SG is sequential, which limits parallelization across measurement steps and the benefit SG can draw from modern GPU hardware. By contrast, C-LBCS is trained via gradient descent on a differentiable cost function and can be fully parallelized across all sub-schemes and all observable terms within each mini-batch, benefiting directly from GPU acceleration.
 The number of sub-schemes of C-LBCS is set to be the same as the number of groups generated in OGM.
Since the average one-shot variance
is defined in the limit of infinitely many measurements, it is hard to estimate it in the exact same way for Derand and SG because the $\{h_j\}$ in these methods cannot be analytically calculated. Here, for a Hamiltonian with $n_H$ terms, we choose to generate $3n_H$ measurements for each method and calculate the average one-shot variance by the measurement scheme that uniformly samples the $3n_H$ measurements.
We note that $3n_H$ is a number much larger than that used in the original works of Derand and SG (1000 shots). Some of the exact numbers of $n_H$ are shown in \autoref{tab:batchsize}.

The result of experiments is shown in \autoref{tab:var-vs-methods}. 
We see that in all cases except for $\mathrm{H_{2}O}$ molecule with BK transformation, C-LBCS outperforms previous methods, which confirms the validity of the composite measurement schemes. Our results agree with the results of the work of SG, though they are not using average one-shot variance as the metrics.

\newcommand{\vartablecaption}{\caption{Table for the average one-shot variance with various measurement methods. C-LBCS with a moderate number of sub-schemes outperforms the other methods in nearly all the systems.
\label{tab:var-vs-methods}}}

\begin{table}[!h]\centering
\begin{tabular}{c|c|cccc}
\toprule 
Molecule & Enc.  & Derand & OGM & SG & C-LBCS\\ 
\hline 
\multirow{2}{*}{$\mathrm{LiH}$(12)}
& JW & 8.26& 7.62& 6.85& \textbf{6.53}\\& BK & 10.68& 7.85& 7.16& \textbf{6.77}\\
\hline 
\multirow{2}{*}{$\mathrm{H_{6}}$(12)}
& JW & 32.64& 30.47& 27.35& \textbf{24.93}\\& BK & 48.25& 35.91& 35.21& \textbf{30.68}\\
\hline 
\multirow{2}{*}{$\mathrm{H_{2}O}$(14)}
& JW & 712& 479& 432& \textbf{430}\\& BK & 891& 518& \textbf{454}& 455\\
\hline 
\multirow{2}{*}{$\mathrm{NH_{3}}$(16)}
& JW & 833& 357& 289& \textbf{287}\\& BK & 1111& 399& 312& \textbf{309}\\
\hline 
\multirow{2}{*}{$\mathrm{N_{2}}$(20)}
& JW & 1571& 1115& 827& \textbf{811}\\& BK & 2002& 1045& 852& \textbf{841}\\
\hline 
\multirow{2}{*}{$\mathrm{C_{2}H_{2}}$(24)}
& JW & 1708& 880& 638& \textbf{580}\\& BK & 2424& 879& 645& \textbf{614}\\
\hline 
\multirow{2}{*}{$\mathrm{C_{2}H_{4}}$(28)}
& JW & 3692& 1481& 1004& \textbf{928}\\& BK & 4908& 1641& 1040& \textbf{1018}\\
\hline 
\multirow{2}{*}{$\mathrm{CO_{2}}$(30)}
& JW & 11655& 3500& 2442& \textbf{2335}\\& BK & 16349& 4169& 2754& \textbf{2677}\\
\hline 
\botrule\end{tabular} 
\vartablecaption
\end{table}

We also study how the average one-shot variance of C-LBCS changes with the number of sub-schemes. The results are shown in \autoref{fig:subscheme-number}, with SG as a reference of variance (red line) and OGM as a reference of the number of sub-schemes (yellow line).
We found that C-LBCS constantly outperforms SG with a number of sub-schemes comparable to the number of OGM groups. \autoref{fig:subscheme-number} also implies that better measurement efficiency can be obtained by increasing the number of sub-schemes. This property of C-LBCS distinguishes it from many previous methods, in which no hyperparameters can be used to trade measurement efficiency with computational resources.

We remark that the absolute variance improvements of C-LBCS over SG may appear modest in some cases. As shown in \autoref{fig:subscheme-number}, the variance of C-LBCS saturates as the number of sub-schemes grows, indicating diminishing returns from simply increasing the number of sub-schemes. Crucially, C-LBCS provides a \emph{systematic} way to study and improve this tradeoff---by increasing the number of sub-schemes, one can progressively reduce the variance until this saturation regime is reached. Greedy schemes such as SG, by contrast, are constructed by fixed heuristics and offer no such handle for principled improvement.

\subsection{Training}
\label{sec:training}
We numerically study how different strategies for parameter optimization affect the convergence of training. We do ablation tests in which \textit{Rescale} or \textit{TTUR} is turned off during training. Other settings of the training are kept the same as adopted by \autoref{tab:var-vs-methods}. All Hamiltonians are encoded with the JW transformation and the result is shown in \autoref{tab:opt-strategy}. We find that keeping both of the strategies provides the best result in most cases. 

\newcommand{\vartablecaptionOpt}{\caption{The average one-shot variance by C-LBCS with different strategies in optimization.
We find that applying the \textit{Rescale} and \textit{TTUR} strategies can significantly improve the result of training. In most cases, applying both of these strategies provides the best result and in a few cases applying \textit{Rescale} alone gives the best result.
\label{tab:opt-strategy}}}

\begin{table}[!h]\centering
\begin{tabular}{c|ccccc}
\toprule 
Molecule  & None & Rescale & TTUR & Both\\ 
\hline 
\multirow{1}{*}{$\mathrm{N_{2}}$(20)}
& 825& \textbf{807}& 815& 811\\
\hline 
\multirow{1}{*}{$\mathrm{C_{2}H_{2}}$(24)}
& 606& 592& 586& \textbf{580}\\
\hline 
\multirow{1}{*}{$\mathrm{C_{2}H_{4}}$(28)}
& 977& 945& 935& \textbf{928}\\
\hline 
\multirow{1}{*}{$\mathrm{CO_{2}}$(30)}
& 2379& \textbf{2335}& 2346& \textbf{2335}\\
\hline 
\multirow{1}{*}{$\mathrm{H_{6}}$(12)}
& 25.67& 25.35& 25.59& \textbf{24.93}\\
\hline 
\multirow{1}{*}{$\mathrm{H_{8}}$(16)}
& 91.52& 88.76& 83.71& \textbf{83.54}\\
\hline 
\multirow{1}{*}{$\mathrm{H_{10}}$(20)}
& 251& 229& 222& \textbf{219}\\
\hline 
\multirow{1}{*}{$\mathrm{H_{12}}$(24)}
& 581& 521& 501& \textbf{496}\\
\hline 
\botrule\end{tabular} 
\vartablecaptionOpt
\end{table}

We then study how different batch sizes affect the training of C-LBCS. We set the batch sizes $n_b$ to be $1/128$, $1/64$ and $1/32$ of the number of terms in the Hamiltonian $n_H$. To make a fair comparison, we adjust the stop criteria proportionally, so that the optimization terminates when the cost function fails to decrease more than $0.1\%$ in the past $1000\times\frac{500}{n_b}$ steps. The Hamiltonians are encoded with the BK transformation. We show in \autoref{tab:batchsize} that the cost function converges to similar values, which implies the training of C-LBCS is not sensitive to batch size for molecular Hamiltonians. Our result indicates that C-LBCS can be applied on much larger molecular Hamiltonians and trained efficiently on GPUs.

We note that the joint optimization of $\vec{r}$ and $\{\theta^k_i\}$ is non-convex and can have local minima in practice. We observe that training can get stuck in such local minima when the sub-scheme parameters are initialized randomly; however, this issue does not arise when the sub-schemes are initialized based on Hamiltonian terms as described above. Furthermore, the stochastic gradient noise introduced by a small batch size appears to help the optimizer escape local minima~\cite{keskar2016large, kleinberg2018alternative}: as seen in \autoref{tab:batchsize}, smaller batch sizes tend to yield slightly lower variance, consistent with this interpretation.

\newcommand{\vartablecaptionBatch}{\caption{The average one-shot variance by C-LBCS with batch sizes to be $1/128$, $1/64$ and $1/32$ of the number of terms in the Hamiltonian ($n_H$). It can be seen that the batch size does not significantly affect the training result of C-LBCS.
\label{tab:batchsize}
}}
\begin{table}[!h]\centering
\begin{tabular}{c|c|cccc}
\toprule 
Molecule & $n_H$ & $1/128$ & $1/64$ & $1/32$\\ 
\hline 
\multirow{1}{*}{$\mathrm{H_{2}O}$(14)}
& 1085 & \textbf{452.73}& 454.39& 454.85\\
\hline 
\multirow{1}{*}{$\mathrm{NH_{3}}$(16)}
& 2936 & \textbf{307.15}& 307.44& 307.56\\
\hline 
\multirow{1}{*}{$\mathrm{N_{2}}$(20)}
& 2238 & \textbf{837.86}& 838.31& 839.16\\
\hline 
\multirow{1}{*}{$\mathrm{C_{2}H_{2}}$(24)}
& 5184 & 611.07& \textbf{610.78}& 611.42\\
\hline 
\multirow{1}{*}{$\mathrm{C_{2}H_{4}}$(28)}
& 8918 & 1014.12& \textbf{1013.51}& 1014.39\\
\hline 
\multirow{1}{*}{$\mathrm{CO_{2}}$(30)}
& 11433 & \textbf{2663.98}& 2666.21& 2678.22\\
\hline 
\botrule\end{tabular} 
\vartablecaptionBatch
\end{table}

\section{Discussion \& Outlook}
\label{section:discussion}

In this work, we proposed a framework for learning measurement schemes from the observable structure, and demonstrated its effectiveness through a scalable and trainable instantiation. C-LBCS provides the parameterization that makes end-to-end optimization feasible. We numerically showed that C-LBCS, trained with gradient rescaling and TTUR motivated by a biconvex structure of the cost function, outperforms previous state-of-the-art methods on all the systems we test given enough sub-schemes. We also found that C-LBCS can be trained by stochastic gradient descent with small batch sizes, enabling the approach to scale to much larger observables.

A key practical advantage of C-LBCS over greedy approaches such as SG~\cite{gresch2023guaranteed} is computational efficiency. Because SG constructs its measurement scheme through a sequential greedy procedure, parallelization across measurement steps is limited, which restricts the benefit SG can draw from GPU acceleration. C-LBCS, by contrast, is trained through batched gradient descent and is fully amenable to GPU parallelism. We additionally present timing results in \appref{section:training-time}, which shows that C-LBCS reaches the final variance achieved by OGM in a fraction of the OGM preprocessing time and benefit from multi-GPU parallelization.
The classical overhead of C-LBCS beyond training is also small. Generating each measurement only takes $O(n_S + n_q)$ time, in which a sub-scheme is sampled from $\vec{r}$ and then a Pauli operator is sampled for each qubit, where $n_S$ is the number of sub-schemes and $n_q$ is the number of qubits. Moreover, the training is a one-time cost: in a typical workflow such as VQE, the trained scheme is reused over the many measurement rounds that follow, so the training time is amortized and is negligible compared to the quantum measurement time in practice. C-LBCS also provides a principled way to trade the number of sub-schemes for measurement efficiency, as discussed in \autoref{sec:var-vs-method}.

We also remark on the assumptions behind C-LBCS and the scenarios where it may be less effective. First, the cost function we adopt is defined in the limit of infinitely many measurements. When the total measurement budget is small relative to the number of terms in the observable, the average one-shot variance may not faithfully reflect the finite-shot estimation error, and a scheme trained with it is no longer guaranteed to be near-optimal for the actual budget. Indeed, \appref{sec:var-vs-shots} shows that on the $\mathrm{CO_2}$ Hamiltonian, SG attains a lower finite-shot variance than C-LBCS when the number of shots is below roughly $3\times 10^4$, with C-LBCS becoming preferable at larger budgets. Second, C-LBCS approximates the optimal measurement distribution by a mixture of product (LBCS-type) distributions. When the optimal distribution is highly structured and cannot be well-approximated by a moderate-size mixture of product distributions, a large number of sub-schemes may be required, and sub-schemes with more complex structure may be more appropriate.

We used the average one-shot variance as our cost function and performance quantifier, which does not involve any information about the state to be measured. This is different from many previous works~\cite{wu2023overlapped, hadfield2022measurements, gresch2023guaranteed} which use the variance concerning a certain state (e.g. the ground state of the Hamiltonian) as the performance quantifier. We chose not to follow this approach because we do not expect the ground state (or its approximation) to be always available, especially when 40 or more qubits are involved. For the same reason, we also adopted a cost function that does not involve the information from a pre-calculated state.
However, in practice, it might be advantageous to use the information of states collected from measurement results to improve the estimation efficiency~\cite{shlosberg2023adaptive}. We leave improving C-LBCS this way as a future question to be investigated.

For more generalization of this work, as we only demonstrated one type of composite measurement scheme, it will be interesting to see whether better measurement schemes can be produced by applying $\operatorname{SampleProd}$ on other types of sub-schemes with more complex structure \cite{hillmich2021decision, akhtar2022scalable}.
Also, as we assumed that only Pauli measurements are allowed considering the implementation hardness, it will be interesting to see how our framework can be generalized to more types of measurements, such as Clifford measurements.

\begin{acknowledgments}
The authors thank Anders G. Frøseth for his generous support. K.N. acknowledges the support of Grant-in-Aid for JSPS Research Fellow 22J01501.
A.A.-G. also acknowledges the generous support of Natural Resources Canada and the Canada 150 Research Chairs program.
\end{acknowledgments}

\section*{Code availability}
The source code used for the numerical simulations is available at \url{https://github.com/aspuru-guzik-group/C-LBCS}.

\bibliography{main}

\appendix
\newpage
\widetext

\section{Evaluation of average one-shot variance for C-LBCS}
\label{sec:h-for-c-lbcs}

In this section, we provide the formula for calculating $\{h^k_j\}$ for C-LBCS schemes, so that the average one-shot variance of it can be calculated by \autoref{equ:cost-fun-decomposed}.

Notice that when a LBCS scheme samples a Pauli string, the sampling of the Pauli operator on each qubit is independent of each other. Therefore, we can simply use the multiplication rule to calculate the probability. Denote the Pauli operator on the $i$-th qubit in the Pauli string $P_j$ by $P_j[i]$ and denote the probability that Pauli operator $P$ is sampled from the distribution $\beta^k_i$ by $\beta^k_i(P)$, we have
\begin{equation}
    h^k_j = \prod_{i=1}^{n_q} p(P_j[i], \beta^k_i),
\end{equation}
where we define
\begin{equation}
p(P_j[i], \beta^k_i)=
\begin{cases}
  \beta^k_i(P_j[i]), & P_j[i] \in \{X,Y,Z\} \\
  1, & P_j[i] = I 
\end{cases}.
\end{equation}

\section{Proof of \autoref{thm:average_cov}}
\label{section:proofLemmaHaar}
With $|\bar{0}\rangle = |0\rangle^{\otimes n_q}$, it holds
\begin{align}
\label{eq:pauliHaarDeriviation}
	\int_{{\rm Haar}} 
	\left(\langle P_j P_{\ell}\rangle - \langle P_j \rangle\langle P_{\ell}\rangle \right) \mathrm{d}|\psi \rangle
	&=
    \int_{\rm Haar}  \mathrm{d} U 
    \left(\langle \bar{0}| U^{\dagger} P_j P_{\ell} U|\bar{0}\rangle
    - \langle \bar{0}|U^{\dagger} P_j U |\bar{0}\rangle \langle \bar{0}| U^{\dagger}P_{\ell} U |\bar{0}\rangle\right).
\end{align}
For calculating the average over the whole Hilbert space with Haar measure, we use the element-wise integration formula \cite{Collins2006}:
\begin{align}
\label{eq:integralTwo}
\int_{\rm Haar} \mathrm{d}U U_{a_1b_1}U_{a_1^{\prime} b_1^{\prime}}^{\ast} &= \frac{1}{N}\delta_{a_1 a_1^{\prime}}\delta_{b_1 b_1^{\prime}}, \\
\label{eq:integralFour}
\int_{\rm Haar} \mathrm{d}U U_{a_{1} b_{1}} U_{a_{2} b_{2}} U_{a_{1}^{\prime} b_{1}^{\prime}}^{*} U_{a_{2}^{\prime} b_{2}^{\prime}}^{*} 
&= \frac{\delta_{a_{1} a_{1}^{\prime}} \delta_{a_{2} a_{2}^{\prime}} \delta_{b_{1} b_{1}^{\prime}} \delta_{b_{2} b_{2}^{\prime}}+\delta_{a_{1} a_{2}^{\prime}} \delta_{a_{2} a_{1}^{\prime}} \delta_{b_{1} b_{2}^{\prime}} \delta_{b_{2} b_{1}^{\prime}}}{N^{2}-1}-
 \frac{\delta_{a_{1} a_{1}^{\prime}} \delta_{a_{2} a_{2}^{\prime}} \delta_{b_{1} b_{2}^{\prime}} \delta_{b_{2} b_{1}^{\prime}}+\delta_{a_{1} a_{2}^{\prime}} \delta_{a_{2} a_{1}^{\prime}} \delta_{b_{1} b_{1}^{\prime}} \delta_{b_{2} b_{2}^{\prime}}}{N\left(N^{2}-1\right)},
\end{align} 
with $N$ as the dimension of the Hilbert space of $U$. 
For the first term in \autoref{eq:pauliHaarDeriviation}, it holds 
\begin{equation}
\label{eq:firstIntegral}
\begin{split}    
\int_{\rm Haar}  {\rm d} U 
    \langle \bar{0}| U^{\dagger} P_j P_{\ell} U|\bar{0}\rangle 
    &= \sum_{a,b}U^{\ast}_{a0} [P_j P_{\ell}]_{ab} U_{b0} \\
    &= \frac{1}{2^{n_q}}{\rm Tr}\left(P_j P_{\ell}\right),
\end{split}
\end{equation}
where we denote $(a,b)$ element of $P_j P_{\ell}$ as $[P_j P_{\ell}]_{ab}$, and we used \autoref{eq:integralTwo} in the second equality. 
For the second term in \autoref{eq:pauliHaarDeriviation}, it holds
\begin{equation}
\label{eq:secondIntegral}
\begin{split}    
\int_{\rm Haar} {\rm d} U \langle \bar{0}|U^{\dagger} P_j U |\bar{0}\rangle \langle \bar{0}| U^{\dagger}P_{\ell} U |\bar{0}\rangle
&= \sum_{a,b,c,d}
U^{\ast}_{a0} [P_j]_{ab} U_{b0}U^{\ast}_{c0} [P_{\ell}]_{cd} U_{d0}\\
&= \frac{1}{2^{2n_q} - 1}\sum_{a,b,c,d}[P_j]_{ab}[P_{\ell}]_{cd}
\left(
\delta_{ba}\delta_{dc} + \delta_{bc}\delta_{da} 
- \frac{1}{2^{n_q}}(\delta_{ba}\delta_{dc} + \delta_{bc}\delta_{da})
\right) \\
&= \frac{1}{2^{2n_q}-1}\left(1 - \frac{1}{2^{n_q}}\right)\left({\rm Tr}(P_j){\rm Tr}(P_{\ell}) + {\rm Tr}(P_j P_{\ell})\right) \\
&= \frac{1}{2^{n_q}(2^{n_q} + 1)} \left({\rm Tr}(P_j){\rm Tr}(P_{\ell}) + {\rm Tr}(P_j P_{\ell})\right),
\end{split}    
\end{equation}   
where in the second equality, we use \autoref{eq:integralFour}. Finally, substituting \autoref{eq:firstIntegral} and \autoref{eq:secondIntegral} into \autoref{eq:pauliHaarDeriviation}, we obtain
\begin{equation}
\begin{split}    
    \int_{\rm Haar} 
	\left(\langle P_j P_{\ell}\rangle - \langle P_j \rangle\langle P_{\ell}\rangle \right) \mathrm{d}|\psi \rangle
	&= \frac{1}{2^{n_q}} {\rm Tr}(P_j P_{\ell}) - \frac{1}{2^{n_q}(2^{n_q} + 1)} \left({\rm Tr}(P_j){\rm Tr}(P_{\ell}) + {\rm Tr}(P_j P_{\ell})\right) \\
    &= \delta_{j{\ell}} \frac{2^{n_q}}{2^{n_q} + 1}, 
\end{split}
\end{equation}
where in the second equality, we used 
\begin{equation}
\label{eq:pauliTrace}
	{\rm Tr}\left(P_j P_{\ell} \right) = 2^{n_q}\delta_{j\ell},~{\rm Tr}\left(P_j\right) = 0, 
\end{equation}
which holds since $P_j, P_{\ell} \in \{X, Y, Z, I\}^{\otimes n_q}$ and $P_j, P_{\ell} \neq I^{\otimes n_q}$ with $X$, $Y$, $Z$ being Pauli operators and $I$ being the identity operator.

\section{Derivation of \autoref{equ:exact-variance}}
\label{sec:var-expression}

To prove
\begin{equation}
    \begin{split}
        \Var(\RV{\expectv{O}}) &= \sum_{j,\ell} a_j a_{\ell} \Cov(\RV{\expectv{P_j}}, \RV{\expectv{P_{\ell}}})
= \sum_{j,\ell} a_j a_{\ell}  \frac{m_{j\ell}}{m_{jj}m_{\ell\ell}} (\expectv{P_j P_{\ell}} - \expectv{P_j}\expectv{P_{\ell}}),
    \end{split}
\end{equation}
It will suffice by showing
\begin{equation}
    \Cov(\RV{\expectv{P_j}}, \RV{\expectv{P_{\ell}}})
= \frac{m_{j\ell}}{m_{jj}m_{\ell\ell}}(\expectv{P_j P_{\ell}} - \expectv{P_j}\expectv{P_{\ell}}).
\end{equation}
Recall that $\RV{\expectv{P_j}}$ and $\RV{\expectv{P_\ell}}$ can be decomposed into the summation
\begin{equation}
    \RV{\expectv{P_j}} := \frac{1}{m_{jj}}\sum_{k,P_j \triangleright Q_k} \mu(P_j, \RV{x}_k),
\end{equation}
which can be substituted into the expression of covariance and yields
\begin{align}
    &\Cov(\RV{\expectv{P_j}}, \RV{\expectv{P_{\ell}}})\\
    =&\Cov(\frac{1}{m_{jj}}\sum_{k,P_j \triangleright Q_k} \mu(P_j, \RV{x}_k), \frac{1}{m_{\ell\ell}}\sum_{k,P_\ell \triangleright Q_k} \mu(P_\ell, \RV{x}_k))\\
    =&\frac{1}{m_{jj}m_{\ell\ell}}\Cov(\sum_{k,P_j \triangleright Q_k} \mu(P_j, \RV{x}_k), \sum_{k,P_\ell \triangleright Q_k} \mu(P_\ell, \RV{x}_k))\\
    =&\frac{1}{m_{jj}m_{\ell\ell}}\sum_{{k_1},P_j \triangleright Q_{k_1}}\sum_{{k_2},P_\ell \triangleright Q_{k_2}}\Cov(\mu(P_j,\RV{x}_{k_1}),\mu(P_\ell, \RV{x}_{k_2})).
\end{align}
Notice that when $k_1\neq k_2$, $\mu(\cdot, \RV{x}_{k_1})$ and $\mu(\cdot, \RV{x}_{k_2})$ are independent of each other and their covariance $\Cov(\mu(\cdot, \RV{x}_{k_1}), \mu(\cdot, \RV{x}_{k_2}))$ is zero. Therefore, The equation can be simplified to
\begin{align}
    &\Cov(\RV{\expectv{P_j}}, \RV{\expectv{P_{\ell}}})\\
    =&\frac{1}{m_{jj}m_{\ell\ell}}
    \sum_{k,P_j \triangleright Q_{k},P_\ell \triangleright Q_{k}}
    \Cov(\mu(P_j,\RV{x}_{k}),\mu(P_\ell, \RV{x}_{k}))\\
    =&\frac{1}{m_{jj}m_{\ell\ell}}\sum_{k,P_j \triangleright Q_{k},P_\ell \triangleright Q_{k}}\Bigg(
    \ExpV{\mu(P_j,\RV{x}_{k})\mu(P_\ell, \RV{x}_{k})}-\ExpV{\mu(P_j,\RV{x}_{k})}\ExpV{\mu(P_\ell, \RV{x}_{k})}\Bigg)\\
\end{align}
By the definition of $\mu(\cdot,\RV{x}_{k})$, it can be deduced that, when  $P_j \triangleright Q_{k},P_\ell \triangleright Q_{k}$, $\ExpV{\mu(P_j,\RV{x}_{k})}$ and  $\ExpV{\mu(P_\ell,\RV{x}_{k})} $ equal to $\expectv{P_j}$ and $\expectv{P_\ell}$ respectively. It can also be derived from the definition that
$\ExpV{\mu(P_j,\RV{x}_{k})\mu(P_\ell, \RV{x}_{k})} = \expectv{P_jP_{\ell}}$. With the definition $m_{j\ell} = \sum_{k,P_j \triangleright Q_{k},P_\ell \triangleright Q_{k}} 1$, we have the final expression
\begin{align}
    \Cov(\RV{\expectv{P_j}}, \RV{\expectv{P_{\ell}}})
    =\frac{m_{j\ell}}{m_{jj}m_{\ell\ell}}(\expectv{P_jP_{\ell}}-\expectv{P_j}\expectv{P_{\ell}}).
\end{align}

\section{Taylor expansion of expectation value}
\label{sec:probability}
Let us first give the Taylor expansion of $f(x)=\frac{1}{x}$ to the first order with the Lagrange remainder:
\begin{equation}
    \frac{1}{x} = \frac{1}{a} - \frac{1}{a^2}(x-a) +\frac{1}{c(x,a)^3}(x-a)^2,
\end{equation}
where $c(x, a)$ is a real number in between $x$ and $a$. 
Setting $a$ to be $\ExpV{\RV{m}_{j}}+\epsilon$ and $x$ to be $\RV{m}_{j}+\epsilon$, we have
\begin{align}
    \frac{M}{\RV{m}_{j}+\epsilon} = \frac{M}{\ExpV{\RV{m}_{j}}+\epsilon} - \frac{M}{(\ExpV{\RV{m}_{j}}+\epsilon)^2}(\RV{m}_{j}-\ExpV{\RV{m}_{j}}) +\frac{M}{c(\RV{m}_{j}+\epsilon, \ExpV{\RV{m}_{j}}+\epsilon)^3}(\RV{m}_{j}-\ExpV{\RV{m}_{j}})^2.
\end{align}
Taking expectation value on both sides, we have
\begin{align}
    \ExpV{\frac{M}{\RV{m}_{j}+\epsilon}} = \frac{M}{\ExpV{\RV{m}_{j}}+\epsilon} +\frac{M}{c(\RV{m}_{j}+\epsilon, \ExpV{\RV{m}_{j}}+\epsilon)^3} \Var(\RV{m}_{j}).
\end{align}
In the following, we will prove that, if $\ExpV{\RV{m}_{j}}\in \Theta(M)$, $\Var(\RV{m}_{j}) \in \Theta(M)$ and $\epsilon = \Theta(M^{5/6})$, we have
\begin{align}
    \lim_{M\rightarrow+\infty}\ExpV{\frac{M}{\RV{m}_{j}+\epsilon}} = \lim_{M\rightarrow+\infty} \frac{M}{\ExpV{\RV{m}_{j}}}.
\end{align}
These assumptions are satisfied if $\RV{m}_{j}$ is a binomial distribution with $M$ experiments.
By using $c(\RV{m}_{j}+\epsilon, \ExpV{\RV{m}_{j}}+\epsilon) \leq M + \epsilon$, we have the lower bound
\begin{align}
    \ExpV{\frac{M}{\RV{m}_{j}+\epsilon}} \geq\frac{M}{\ExpV{\RV{m}_{j}}+\epsilon} +\frac{M}{(M+\epsilon)^3}\Var(\RV{m}_{j}).
\end{align}
In the limit that $M\rightarrow+\infty$, we have 
\begin{align}
    \lim_{M\rightarrow+\infty}\ExpV{\frac{M}{\RV{m}_{j}+\epsilon}} \geq \lim_{M\rightarrow+\infty} \Big( \frac{1}{\ExpV{\RV{m}_{j}}/M+ \Theta(M^{5/6})/M} + \frac{M}{\Theta(M)^3} \Theta(M) \Big) = \lim_{M\rightarrow+\infty} \frac{M}{\ExpV{\RV{m}_{j}}}.
\end{align}
For the other side, by using $c(\RV{m}_{j}+\epsilon, \ExpV{\RV{m}_{j}}+\epsilon) \geq \epsilon$, we have
\begin{align}
    \ExpV{\frac{M}{\RV{m}_{j}+\epsilon}} \leq \frac{M}{\ExpV{\RV{m}_{j}}+\epsilon} +\frac{M}{\epsilon^3}\Var(\RV{m}_{j}).
\end{align}
In the limit that $M\rightarrow+\infty$, we have
\begin{align}
    \lim_{M\rightarrow+\infty}\ExpV{\frac{M}{\RV{m}_{j}+\epsilon}} \leq \lim_{M\rightarrow+\infty} \frac{M}{\ExpV{\RV{m}_{j}}} +\lim_{M\rightarrow+\infty}\frac{\Theta(M^2)}{\Theta(M^{5/6})^3}= \lim_{M\rightarrow+\infty} \frac{M}{\ExpV{\RV{m}_{j}}}.
\end{align}
Therefore, by bounds from both sides, we can see that 
\begin{align}
    \lim_{M\rightarrow+\infty}\ExpV{\frac{M}{\RV{m}_{j}+\epsilon}} = \lim_{M\rightarrow+\infty} \frac{M}{\ExpV{\RV{m}_{j}}}.
\end{align}

\section{Detail about training}
\label{section:detailTraning}
All the optimizations in this work are done by PyTorch~\cite{paszke2019pytorch} with the Adam optimizer. To parameterize every probability vector $\vec{r}$,  which satisfies $r_i>0$ and $\sum r_i=1$, we use a vector $\vec{y}$ of real numbers and set $\vec{r}=\operatorname{Normalize}(\operatorname{SoftPlus}(\vec{\theta}))$. Here, $\operatorname{SoftPlus}$ is applied element-wise to $\vec{\theta}$ as
\begin{equation}
    \operatorname{SoftPlus}(\theta_i) = \frac{1}{\beta} \log(1+\exp(\beta \times \theta_i)).
\end{equation}
$\operatorname{Normalize}$ is defined as
\begin{equation}
    \operatorname{Normalize}(\vec{z}) = \frac{\vec{z}}{\sum_i z_i}.
\end{equation}

\section{Biconvex structure of the cost function}
\label{section:appendix-biconvex}

In this appendix, we formally state and prove that the cost function $V$ in \autoref{equ:cost-fun-decomposed} is biconvex in the variables $(\vec{r}, \{h^k_j\})$, and we clarify the scope of this result: the convexity does \emph{not} carry over to the sub-scheme parameters $\{\theta^k_i\}$ that are optimized in practice.

Throughout this appendix, we regard
\begin{equation}
\label{eq:V-two-variables}
    V(\vec{r}, \vec{h}) = \frac{2^{n_q}}{2^{n_q}+1} \sum_j \frac{a_j^2}{g_j(\vec{r}, \vec{h})},
    \qquad
    g_j(\vec{r}, \vec{h}) := \sum_{k} r_k h^k_j,
\end{equation}
as a function of the probabilities $\vec{r}$ and the collection of coverage probabilities $\vec{h} = \{h^k_j\}$, treated as independent variables. Let $\Delta := \{\vec{r} \in \mathbb{R}^{n_S} : r_k \geq 0, \sum_k r_k = 1\}$ denote the probability simplex, with $n_S$ the number of sub-schemes.

\begin{proposition}[Biconvexity of $V$]
\label{prop:biconvex}
(i) For every fixed $\vec{r} \in \Delta$, the function $\vec{h} \mapsto V(\vec{r}, \vec{h})$ is convex on the convex set
$D_{\vec{h}}(\vec{r}) := \{\vec{h} : h^k_j \in [0,1] \text{ for all } k, j, \text{ and } g_j(\vec{r}, \vec{h}) > 0 \text{ for all } j\}$.
(ii) For every fixed $\vec{h}$ with $h^k_j \in [0,1]$, the function $\vec{r} \mapsto V(\vec{r}, \vec{h})$ is convex on the convex set
$D_{\vec{r}}(\vec{h}) := \{\vec{r} \in \Delta : g_j(\vec{r}, \vec{h}) > 0 \text{ for all } j\}$.
In particular, $V$ is biconvex.
\end{proposition}

\begin{proof}
We first verify that the domains are convex. The box $\{\vec{h} : h^k_j \in [0,1]\}$ and the simplex $\Delta$ are convex sets. For fixed $\vec{r}$, each map $\vec{h} \mapsto g_j(\vec{r}, \vec{h})$ is affine in $\vec{h}$; for fixed $\vec{h}$, each map $\vec{r} \mapsto g_j(\vec{r}, \vec{h})$ is affine in $\vec{r}$. Hence each set $\{g_j > 0\}$ is an open half-space in the free variables, and $D_{\vec{h}}(\vec{r})$ and $D_{\vec{r}}(\vec{h})$, being intersections of convex sets, are convex.

We will use the following elementary facts. (a) The function $f(x) = 1/x$ is convex on $(0, +\infty)$, since $f''(x) = 2/x^3 > 0$ there. (b) The composition of a convex function with an affine map is convex. Here, a map $g$ is \emph{affine} if it has the form $g(\vec{x}) = \vec{c}^\top \vec{x} + d$ for some fixed vector $\vec{c}$ and scalar $d$, i.e., a linear function plus a constant; each map $g_j(\vec{r}, \vec{h}) = \sum_k r_k h^k_j$ appearing below is affine in $\vec{h}$ for fixed $\vec{r}$ (with $\vec{c} = \vec{r}$, $d=0$), and affine in $\vec{r}$ for fixed $\vec{h}$ (with $\vec{c} = \vec{h}$, $d=0$). Formally, if $g$ is affine, then for any $\vec{x}, \vec{y}$ in a convex set where $f \circ g$ is defined and any $t \in [0, 1]$,
\begin{equation}
    f(g(t\vec{x} + (1-t)\vec{y})) = f(t\, g(\vec{x}) + (1-t)\, g(\vec{y})) \leq t\, f(g(\vec{x})) + (1-t)\, f(g(\vec{y})).
\end{equation}
(c) A summation of convex functions with non-negative coefficients is convex.

To prove (i), fix $\vec{r} \in \Delta$. Each $\vec{h} \mapsto g_j(\vec{r}, \vec{h})$ is affine and strictly positive on $D_{\vec{h}}(\vec{r})$, so by (a) and (b), $\vec{h} \mapsto 1/g_j(\vec{r}, \vec{h})$ is convex on $D_{\vec{h}}(\vec{r})$ for every $j$. By (c), $V(\vec{r}, \cdot)$ in \autoref{eq:V-two-variables} is convex on $D_{\vec{h}}(\vec{r})$, since $a_j^2 \geq 0$ and the prefactor is positive. The proof of (ii) is identical with the roles of $\vec{r}$ and $\vec{h}$ exchanged: for fixed $\vec{h}$, each $\vec{r} \mapsto g_j(\vec{r}, \vec{h})$ is affine and strictly positive on $D_{\vec{r}}(\vec{h})$, so every term $a_j^2 / g_j(\vec{r}, \vec{h})$, and therefore $V(\cdot, \vec{h})$, is convex on $D_{\vec{r}}(\vec{h})$.
\end{proof}

In practice, $\vec{r}$ is represented by the $\operatorname{SoftPlus}$-$\operatorname{Normalize}$ parameterization described in the last section. This parameterization realizes exactly the open probability simplex: $\operatorname{SoftPlus}$ maps $\mathbb{R}^{n_S}$ onto the positive orthant, $\operatorname{Normalize}$ maps the positive orthant onto the open simplex, and any $\vec{r}$ with strictly positive entries is realized by, e.g., $\theta_k = \operatorname{SoftPlus}^{-1}(r_k)$. Optimizing over the raw parameters therefore searches the full domain of the convex subproblem over $\vec{r}$. However, since neither $\operatorname{SoftPlus}$ nor $\operatorname{Normalize}$ is affine, the convexity established in \autoref{prop:biconvex} is not guaranteed to be preserved as a function of the raw parameters $\vec{\theta}$.

We caution that this biconvexity is a statement about the variables $(\vec{r}, \{h^k_j\})$ and does not extend to the parameters that are actually optimized in C-LBCS. Each coverage probability is a product over qubits, $h^k_j = \prod_i p(P_j[i], \beta^k_i)$ (\appref{sec:h-for-c-lbcs}), and this nonlinear map does not preserve convexity, so $V$ is in general not convex in the per-qubit distributions $\{\beta^k_i\}$ (or in the raw parameters $\{\theta^k_i\}$ behind the $\operatorname{SoftPlus}$-$\operatorname{Normalize}$ parameterization) even when $\vec{r}$ is fixed.

Therefore, the convex subproblems guaranteed by the biconvex structure are those over $\vec{r}$ (with $\{h^k_j\}$ fixed) and over $\{h^k_j\}$ (with $\vec{r}$ fixed), while the subproblems over the raw parameters solved in practice are non-convex in general. The biconvex structure nevertheless indicates that the cost function is considerably more benign than generic non-convex objectives, and it motivates the TTUR strategy of updating $\vec{r}$ and $\{\theta^k_i\}$ with different learning rates.

\section{Time used for training}
\label{section:training-time}

We present the running time of OGM and C-LBCS classical pre-processing for the JW-encoded Hamiltonians in \autoref{tab:var-vs-methods}. The GPU parts of the calculation are measured on one NVIDIA GeForce RTX 2070 graphics card, while the CPU parts are measured on an Intel Xeon Gold 6248R CPU @ 3.00\,GHz. For OGM, the grouping construction is performed on the CPU and the subsequent mixing-weight optimization is performed on the GPU.

\newcommand{\timetablecaption}{\caption{
Wall-clock time (seconds) for OGM and C-LBCS classical pre-processing on JW-encoded molecular Hamiltonians, measured on the hardware described in the text. C-LBCS$^*$ is the time for C-LBCS training to first reach the final variance of OGM; C-LBCS is the total training time.
\label{tab:training-time}
}}

\begin{table}[!h]\centering
\begin{tabular}{c|ccc}
\toprule
Molecule & OGM (s) & C-LBCS$^*$ (s) & C-LBCS (s)\\
\hline
$\mathrm{LiH}$(12) & 31.5 & 4.2 & 37.4\\
\hline
$\mathrm{H_{6}}$(12) & 8.6 & 6.0 & 46.1\\
\hline
$\mathrm{H_{2}O}$(14) & 23.8 & 3.3 & 35.5\\
\hline
$\mathrm{NH_{3}}$(16) & 40.6 & 5.4 & 81.1\\
\hline
$\mathrm{N_{2}}$(20) & 30.1 & 5.7 & 84.8\\
\hline
$\mathrm{C_{2}H_{2}}$(24) & 69.1 & 11.4 & 142\\
\hline
$\mathrm{C_{2}H_{4}}$(28) & 181 & 25.4 & 267\\
\hline
$\mathrm{CO_{2}}$(30) & 286 & 43.2 & 378\\
\hline
\botrule\end{tabular}
\timetablecaption
\end{table}

We also measure the \emph{shot rate} of ShadowGrouping (SG), defined as the number of measurement shots it can generate per second.
SG constructs each shot by a greedy qubit-wise-covering procedure that iterates over all Hamiltonian terms; consequently, the per-shot cost scales with the number of terms $n_H$, and the shot rate decreases as the Hamiltonian grows.
This greedy dependence also makes SG difficult to parallelize across the sequential choices within one shot, although lower-level scoring operations can still be parallelized.
We measure the shot rate by timing exactly 1000 shots after a numba JIT warm-up, on one core of the same CPU.
The results are shown in \autoref{tab:sg-shot-rate}.

\newcommand{\sgratecaption}{\caption{%
Shot rate (shots per second) of ShadowGrouping on JW-encoded molecular Hamiltonians, measured on one CPU core.
The shot rate decreases with the number of Hamiltonian terms $n_H$ because each shot requires $O(n_H)$ work to greedily select the next measurement.
\label{tab:sg-shot-rate}}}

\begin{table}[!h]\centering
\begin{tabular}{c|c|c}
\toprule
Molecule & $n_H$ & SG shot rate (shots/s)\\
\hline
$\mathrm{LiH}$(12)        &   630 & 12006\\
\hline
$\mathrm{H_{6}}$(12)      &   918 &  8095\\
\hline
$\mathrm{H_{2}O}$(14)     &  1085 &  7269\\
\hline
$\mathrm{N_{2}}$(20)      &  2238 &  3788\\
\hline
$\mathrm{NH_{3}}$(16)     &  2936 &  3046\\
\hline
$\mathrm{C_{2}H_{2}}$(24) &  5184 &  1570\\
\hline
$\mathrm{C_{2}H_{4}}$(28) &  8918 &   931\\
\hline
$\mathrm{CO_{2}}$(30)     & 11433 &   836\\
\hline
\botrule\end{tabular}
\sgratecaption
\end{table}

This decrease in SG shot rate can become a serious practical issue at larger scales. Unlike one-time preprocessing, measurement generation is performed throughout the sampling process; if the Hamiltonian has many terms, the classical time required to choose each SG measurement can become a bottleneck for high-throughput experiments requiring many shots. The greedy structure further limits how much this bottleneck can be relieved by parallel hardware, because each shot is built through dependent choices rather than a single batched sampling operation. For future practical regimes, such as molecular Hamiltonians with roughly $70$ qubits and $10^6$ terms, this sequential per-shot cost is expected to be substantially more severe than in the $30$-qubit examples tested here. C-LBCS has a different scaling profile: after training, each measurement is sampled directly from the learned mixture without a greedy pass over all Hamiltonian terms. Its heavier cost is the one-time training stage, which consists mainly of batched tensor operations and can be mitigated by multi-GPU parallelization, as we now demonstrate.

We benchmark data-parallel C-LBCS training on $\mathrm{CO_2}$ (30 qubits, JW, 5819 sub-schemes) using 1 to 3 NVIDIA GeForce RTX 2070 GPUs on a single node, via PyTorch DistributedDataParallel with data sharding across GPUs.
In each run, the Hamiltonian terms are partitioned disjointly across GPUs; each GPU computes gradients on its local shard and gradients are synchronized via NCCL all-reduce before the optimizer step.
We report the average per-epoch wall-clock time in \autoref{tab:ddp-scaling}.
The speedup is close to linear, reaching $2.79\times$ with 3 GPUs.

\newcommand{\ddpscalingcaption}{\caption{%
Per-epoch wall-clock time and speedup for data-parallel C-LBCS training on $\mathrm{CO_2}$ (30 qubits, JW, 5819 sub-schemes) with 1--3 NVIDIA GeForce RTX 2070 GPUs on a single node (PyTorch DDP, NCCL all-reduce).
The per-epoch time decreases close to linearly, reaching a $2.79\times$ speedup with 3 GPUs.
\label{tab:ddp-scaling}}}

\begin{table}[!h]\centering
\begin{tabular}{c|c|c}
\toprule
GPUs & Epoch time (s) & Speedup\\
\hline
1 & 0.391 & $1.00\times$\\
\hline
2 & 0.207 & $1.89\times$\\
\hline
3 & 0.140 & $2.79\times$\\
\hline
\botrule\end{tabular}
\ddpscalingcaption
\end{table}

\section{Variance vs.\ number of shots: C-LBCS and ShadowGrouping}
\label{sec:var-vs-shots}

\begin{figure}[h]
    \centering
    \includegraphics[width=0.45\linewidth]{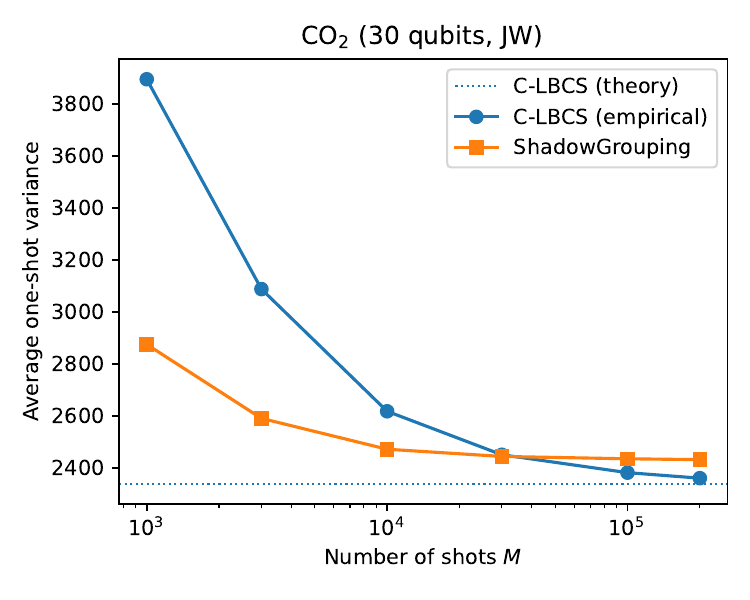}
    \caption{Average one-shot variance vs.\ number of shots $M$ for $\mathrm{CO_2}$ (30 qubits, JW). C-LBCS (empirical) uses $M$ i.i.d.\ shots; C-LBCS (theory) uses analytical $h_j$.}
    \label{fig:var-vs-shots}
\end{figure}

\autoref{fig:var-vs-shots} shows the average one-shot variance as a function of the number of measurement shots $M$ for both C-LBCS and SG on the $\mathrm{CO_2}$ Hamiltonian (30 qubits, JW encoding, 11433 terms).
The C-LBCS scheme used in this figure consists of $5819$ sub-schemes, matching the number of groups produced by OGM for this Hamiltonian, consistent with the setting adopted throughout \autoref{sec:var-vs-method}.

For C-LBCS, the empirical average one-shot variance is computed from $M$ i.i.d.\ shots sampled from the trained distribution.
The same finite-shot convention for estimating the average one-shot variance is used for both methods.
The C-LBCS theory line uses the analytical coverage probabilities $h_j$ computed directly from the trained sub-schemes (see \appref{sec:h-for-c-lbcs}), which serves as the $M\to\infty$ limit of the empirical curve.

Three features of \autoref{fig:var-vs-shots} are worth noting.
First, the empirical C-LBCS average one-shot variance converges to the analytical value as $M$ grows, confirming that the average one-shot variance used as our cost function correctly describes the asymptotic behavior of the trained scheme.
Second, in the small-budget regime ($M \lesssim 3\times 10^4$ for this Hamiltonian), SG attains a lower empirical average one-shot variance than C-LBCS. This is expected from the construction of the two methods: SG builds its measurement list greedily and deterministically, covering the high-weight terms as early as possible, whereas C-LBCS samples measurements i.i.d.\ from the learned mixture, which leaves a larger fraction of terms uncovered when $M$ is small. This concretely illustrates the limitation stated in \autoref{section:discussion}: the average one-shot variance is an asymptotic quantifier, and a scheme trained with it is not guaranteed to be preferable at small measurement budgets, where methods such as SG remain the better choice.
Third, beyond the crossover point, C-LBCS achieves a lower empirical average one-shot variance than SG and approaches its analytical value, consistent with the asymptotic comparison in \autoref{tab:var-vs-methods}.

\section{Handling uncovered terms when computing the average one-shot variance}
\label{sec:uncovered-terms}

When the average one-shot variance $V = \sum_j \frac{a_j^2}{h_j}\frac{2^{n_q}}{2^{n_q}+1}$ is evaluated empirically from a finite set of $M$ shots, some Hamiltonian terms $P_j$ may not be covered by any of the $M$ sampled measurements.
For such \emph{uncovered} terms, the empirical coverage probability $h_j = 0$, rendering the formula ill-defined.
In this section we describe the completion rule used for the empirical curves in \appref{sec:var-vs-shots} and the main text.
Throughout this appendix, every quantity denoted $V$ (with subscripts or arguments) is an average one-shot variance or an empirical estimate thereof, in the sense of the Definition in \autoref{section:optimization}.

\subsection*{Setup}

Let $\mathcal{C}$ and $\mathcal{U}$ denote the index sets of covered and uncovered terms, respectively.
For the covered terms, the empirical distribution $D_1$ yields coverage probabilities $h_j > 0$ for $j\in\mathcal{C}$, and the partial average one-shot variance contribution is
\begin{equation}
    V_1 = \frac{2^{n_q}}{2^{n_q}+1}\sum_{j\in\mathcal{C}} \frac{a_j^2}{h_j}.
\end{equation}
For the uncovered terms, we use an $\ell_1$ fallback contribution corresponding to sampling the uncovered Pauli terms with probabilities proportional to $|a_j|$. Denote this fallback distribution by $D_2$; it is the simple measurement scheme that samples a Pauli measurement covering a single uncovered term $j\in\mathcal{U}$ with probability
\begin{equation}
    h^{(2)}_j = \frac{|a_j|}{\sum_{j'\in\mathcal{U}}|a_{j'}|}, \quad j\in\mathcal{U},
\end{equation}
so that $h^{(2)}_j$ is the coverage probability of $D_2$ for term $j$. This gives the usual $\ell_1$ average one-shot variance contribution for the uncovered sub-observable $\sum_{j\in\mathcal{U}}a_j P_j$,
\begin{equation}
    V_2 = \frac{2^{n_q}}{2^{n_q}+1}\sum_{j\in\mathcal{U}} \frac{a_j^2}{h^{(2)}_j}
        = \frac{2^{n_q}}{2^{n_q}+1}\Bigl(\sum_{j\in\mathcal{U}}|a_j|\Bigr)^2.
\end{equation}

\subsection*{Completion rule used in the finite-shot plot}

Neither $D_1$ nor $D_2$ alone is a valid scheme for the whole observable: $D_1$ never covers $\mathcal{U}$, and $D_2$ never covers $\mathcal{C}$. A valid single scheme is obtained by forming the normalized mixture $D_\alpha = \alpha_1 D_1 + \alpha_2 D_2$ with $\alpha_1,\alpha_2>0$ and $\alpha_1+\alpha_2=1$, under which the effective coverage probabilities become
\begin{equation}
    h_j^{(\alpha)} =
    \begin{cases}
        \alpha_1\, h_j, & j\in\mathcal{C},\\[4pt]
        \alpha_2\, h^{(2)}_j, & j\in\mathcal{U},
    \end{cases}
\end{equation}
in the idealized decomposition where the fallback contribution is assigned only to $\mathcal{U}$. The corresponding average one-shot variance is
\begin{equation}
    V(\alpha_1) = \frac{V_1}{\alpha_1} + \frac{V_2}{\alpha_2} = \frac{V_1}{\alpha_1} + \frac{V_2}{1-\alpha_1}.
\end{equation}
Simply adding $V_1+V_2$, as if $\alpha_1=\alpha_2=1$, is \emph{not} the average one-shot variance of any valid scheme, since a valid mixture requires $\alpha_1,\alpha_2\in(0,1)$ with $\alpha_1+\alpha_2=1$; because $\alpha_1<1$ and $\alpha_2<1$, we have $V_1/\alpha_1 > V_1$ and $V_2/\alpha_2 > V_2$ termwise, so $V(\alpha_1) > V_1+V_2$ for every admissible $\alpha_1$. The naive sum therefore underestimates the average one-shot variance achievable by any properly normalized combination of $D_1$ and $D_2$.

We instead report the average one-shot variance of the optimally mixed scheme. To minimize $V(\alpha_1) = V_1/\alpha_1 + V_2/(1-\alpha_1)$ over $\alpha_1\in(0,1)$, we differentiate with respect to $\alpha_1$,
\begin{equation}
    \frac{\mathrm{d}V}{\mathrm{d}\alpha_1} = -\frac{V_1}{\alpha_1^2} + \frac{V_2}{(1-\alpha_1)^2},
\end{equation}
and set the derivative to zero, which gives the first-order condition
\begin{equation}
    \frac{V_1}{\alpha_1^2} = \frac{V_2}{(1-\alpha_1)^2}
    \implies
    \frac{\sqrt{V_1}}{\alpha_1} = \frac{\sqrt{V_2}}{1-\alpha_1},
\end{equation}
where the second implication takes the (positive) square root of both sides, valid since $V_1,V_2>0$ and $\alpha_1,1-\alpha_1\in(0,1)$. Cross-multiplying, $\sqrt{V_1}(1-\alpha_1) = \sqrt{V_2}\,\alpha_1$, so $\sqrt{V_1} = \alpha_1(\sqrt{V_1}+\sqrt{V_2})$, from which the optimal mixing coefficients follow:
\begin{align}
\label{eq:optimal-mixing}
    \alpha_1^* &= \frac{1}{1+\sqrt{V_2/V_1}}, &
    \alpha_2^* &= \frac{\sqrt{V_2/V_1}}{1+\sqrt{V_2/V_1}} = 1 - \alpha_1^*.
\end{align}
This critical point is indeed a minimum: the second derivative
\begin{equation}
    \frac{\mathrm{d}^2V}{\mathrm{d}\alpha_1^2} = \frac{2V_1}{\alpha_1^3} + \frac{2V_2}{(1-\alpha_1)^3}
\end{equation}
is strictly positive for every $\alpha_1\in(0,1)$ (since $V_1,V_2>0$), so $V(\alpha_1)$ is strictly convex on $(0,1)$ and $\alpha_1^*$ is its unique global minimizer.
Substituting back, the minimum average one-shot variance achieved by the mixed scheme is
\begin{equation}
    V^* = V(\alpha_1^*) = \bigl(\sqrt{V_1}+\sqrt{V_2}\bigr)^2.
\end{equation}
For the finite-shot curves in \appref{sec:var-vs-shots}, we report this optimally mixed value as the completed average one-shot variance estimate,
\begin{equation}
    V_{\mathrm{completed}} := V^* = \bigl(\sqrt{V_1}+\sqrt{V_2}\bigr)^2.
\end{equation}
This convention is applied to all the reported SG variances. C-LBCS variances do not need correction in our experiments.

\subsection*{Implementation}

In practice, the procedure applied to all the variance of the SG method, which produces uncovered terms, is as follows.
Given $M$ sampled shots with coverage probabilities $\{h_j\}$:
\begin{enumerate}
    \item Identify the uncovered significant terms: $\mathcal{U} = \{j : h_j = 0,\; |a_j| > 0\}$.
    \item If $\mathcal{U}=\emptyset$, use $\{h_j\}$ directly.
    \item Otherwise, compute $V_1$ from the covered terms and $V_2 = \frac{2^{n_q}}{2^{n_q}+1}(\sum_{j\in\mathcal{U}}|a_j|)^2$.
    \item Report the completed average one-shot variance estimate $V_{\mathrm{completed}} = \bigl(\sqrt{V_1}+\sqrt{V_2}\bigr)^2$.
\end{enumerate}
This procedure ensures that the plotted empirical average one-shot variance estimate is always finite and that uncovered terms are treated by the same rule for both methods.

\end{document}

%% file: macro_style.tex
\usepackage[ruled,vlined]{algorithm2e}

\newtheorem{theorem}{Theorem}
\newtheorem{lemma}{Lemma}

\newtheorem{definition}{Definition}
\newtheorem{proposition}{Proposition}

\usepackage{color}

\labelformat{figure}{#1}
\labelformat{algorithm}{#1}
\labelformat{theorem}{#1}
\labelformat{proposition}{#1}
\labelformat{equation}{(#1)}

\newcommand{\appref}[1]{\hyperref[#1]{Appendix~\ref*{#1}}}

%% file: macro_quantum.tex
\newcommand{\bra}[1]{\langle#1|}

\newcommand{\ket}[1]{|#1\rangle}

\newcommand{\expectv}[1]{\langle#1\rangle}

%% file: macro_math.tex
\newcommand{\Exp}{\mathbb{E}}
\newcommand{\ExpV}[1]{\mathbb{E}[#1]}
\newcommand{\Var}{\operatorname{Var}}
\newcommand{\Cov}{\operatorname{Cov}}